\documentclass[a4paper,fleqn]{cas-sc}

\usepackage[authoryear]{natbib}

\ExplSyntaxOn \cs_gset:Npn \__first_footerline: { \group_begin: \small \sffamily \__short_authors: \group_end: } \ExplSyntaxOff

\def\tsc#1{\csdef{#1}{\textsc{\lowercase{#1}}\xspace}}
\tsc{WGM}
\tsc{QE}

\usepackage{lineno}
\modulolinenumbers[5]
\UseRawInputEncoding

\usepackage[american]{babel}

\usepackage{times}
\usepackage{helvet}
\usepackage{latexsym}
\usepackage{comment}

\usepackage{bm}
\usepackage{amsfonts}
\usepackage{amssymb}
\usepackage{amsmath}
\usepackage{booktabs}

\usepackage[title]{appendix}

\usepackage[dvips]{epsfig}
\usepackage[dvips]{graphicx}
\usepackage{psfrag}
\usepackage{subcaption}

\usepackage{xcolor}

\usepackage{float}

\usepackage{caption}

\usepackage[noabbrev]{cleveref}

\usepackage{epic,eepic}

\newcommand{\term}[1]{\textsc{#1}}

\newcommand{\ie}{\textit{i.e.}}

\newcommand{\transpose}[1]{\ensuremath{{#1}^{\text T}}}
\newcommand{\inverse}[1]{\ensuremath{{#1}^{-1}}}
\newcommand{\invtranspose}[1]{\ensuremath{{#1}^{\text{-T}}}}
\newcommand{\sign}[1]{\ensuremath{\operatorname{sgn}\left({#1}\right)}}

\newcommand{\Grad}[1][]{\ensuremath{\nabla{#1}}}

\newcommand{\Divergence}[1][]{\ensuremath{\operatorname{Div}{#1}}}

\newcommand{\inc}[1]{\ensuremath{\text d{#1}}}
\newcommand{\abs}[1]{\ensuremath{\left|{#1}\right|}}
\newcommand{\norm}[2][]{\ensuremath{\left|\left|{#2}\right|\right|\if\relax\detokenize{#1}\relax\else _{#1}\fi}}

\newcommand{\domain}[1]{\ensuremath{\mathcal{#1}}}
\newcommand{\tnsrfour}[1]{\ensuremath{\mathbb{#1}}}
\newcommand{\tnsr}[1]{\ensuremath{\mathbf{#1}}}

\newcommand{\vctr}[1]{\ensuremath{\mathbf{#1}}}
\newcommand{\vctrgreek}[1]{\ensuremath{\bm{#1}}}

\newcommand{\eyetwo}{\ensuremath{\tnsr I}}

\newcommand{\defmap}{\ensuremath{\vctrgreek{\chi}}}
\newcommand{\fPK}{\ensuremath{\tnsr P}}

\newcommand{\sPK}{\ensuremath{\tnsr S}}
\newcommand{\F}[1][]{\ensuremath{\tnsr F^{#1}}}

\newcommand{\Fp}[1][]{\ensuremath{\tnsr F_\text{p}^{#1}}}
\newcommand{\Fpdot}[1][]{\ensuremath{\dot{\tnsr F}_\text{p}^{#1}}}
\newcommand{\Fe}[1][]{\ensuremath{\tnsr F_\text{e}^{#1}}}

\newcommand{\Lp}{\ensuremath{\tnsr L_\text{p}}}
\newcommand{\Ee}{\ensuremath{\tnsr E_\text{e}}}

\newcommand{\cs}[2][]{\ensuremath{\sigma^{(#1)}_ \text{#2}}}

\newcommand{\ese}[1][]{\mathcal{W}_e}
\newcommand{\esec}[1][]{{\mathcal{W}}^{crit}_e}

\newcommand{\X}[1][]{\ensuremath{\vctr X}}
\newcommand{\x}[1][]{\ensuremath{\vctr x}}
\newcommand{\disp}[1][]{\ensuremath{\vctr u}}

\newcommand{\dX}[1][]{\ensuremath{\vctr{dX}}}
\newcommand{\dx}[1][]{\ensuremath{\vctr{dx}}}
\newcommand{\dy}[1][]{\ensuremath{\vctr{dy}}}
\newcommand{\du}[1][]{\ensuremath{\vctr{du}}}
\newcommand{\dv}[1][]{\ensuremath{\vctr{dv}}}

\newcommand{\vecx}[1][]{\ensuremath{\vctr{x}}}
\newcommand{\vecX}[1][]{\ensuremath{\vctr{X}}}
\newcommand{\vecy}[1][]{\ensuremath{\vctr{y}}}
\newcommand{\vecu}[1][]{\ensuremath{\vctr{u}}}
\newcommand{\vecv}[1][]{\ensuremath{\vctr{v}}}

\newcommand{\invF}[1][]{\ensuremath{\inverse{\F}}}
\newcommand{\tranF}[1][]{\ensuremath{\transpose{\F}}}
\newcommand{\invtranF}[1][]{\ensuremath{\invtranspose{\F}}}

\newcommand{\pl}{\textsubscript{p}}
\newcommand{\da}{\textsubscript{d}}
\newcommand{\el}{\textsubscript{e}}

\newcommand{\al}{\textsuperscript{\alpha}}

\newcommand{\vecnda}[1][]{\ensuremath{\hat{\vctr{n}}\da{}\al}}

\newcommand{\vecn}[1][]{\ensuremath{\hat{\vctr{n}}}}
\newcommand{\vecd}[1][]{\ensuremath{\hat{\vctr{d}}}}
\newcommand{\vect}[1][]{\ensuremath{\hat{\vctr{t}}}}

\newcommand{\vecnp}[1][]{\ensuremath{\hat{\vctr{n}}\pl}}
\newcommand{\vecdp}[1][]{\ensuremath{\hat{\vctr{d}}\pl}}
\newcommand{\vectp}[1][]{\ensuremath{\hat{\vctr{t}}\pl}}

\newcommand{\vecnd}[1][]{\ensuremath{\hat{\vctr{n}}\da}}
\newcommand{\vecdd}[1][]{\ensuremath{\hat{\vctr{d}}\da}}
\newcommand{\vectd}[1][]{\ensuremath{\hat{\vctr{t}}\da}}

\newcommand{\vecne}[1][]{\ensuremath{\hat{\vctr{n}}\el}}
\newcommand{\vecde}[1][]{\ensuremath{\hat{\vctr{d}}\el}}
\newcommand{\vecte}[1][]{\ensuremath{\hat{\vctr{t}}\el}}

\newcommand{\Fd}[1][]{\ensuremath{\tnsr F\da}}

\newcommand{\invFe}[1][]{\ensuremath{\inverse{\Fe}}}
\newcommand{\tranFe}[1][]{\ensuremath{\transpose{\Fe}}}
\newcommand{\invtranFe}[1][]{\ensuremath{\invtranspose{\Fe}}}

\newcommand{\invFp}[1][]{\ensuremath{\inverse{\Fp}}}
\newcommand{\tranFp}[1][]{\ensuremath{\transpose{\Fp}}}
\newcommand{\invtranFp}[1][]{\ensuremath{\invtranspose{\Fp}}}

\newcommand{\invFd}[1][]{\ensuremath{\inverse{\Fd}}}
\newcommand{\tranFd}[1][]{\ensuremath{\transpose{\Fd}}}
\newcommand{\invtranFd}[1][]{\ensuremath{\invtranspose{\Fd}}}

\newcommand{\Fedot}[1][]{\ensuremath{\dot{\Fe}}}
\newcommand{\Fddot}[1][]{\ensuremath{\dot{\Fd}}}

\newcommand{\svsc}[1][]{\ensuremath{\xi}}
\newcommand{\svdsc}[1][]{\ensuremath{\xi\da}}
\newcommand{\svpsc}[1][]{\ensuremath{\xi\pl}}

\newcommand{\sv}[1][]{\ensuremath{\vctrgreek \xi}}
\newcommand{\svd}[1][]{\ensuremath{\sv\da}}
\newcommand{\svp}[1][]{\ensuremath{\sv\pl}}

\newcommand{\dotsv}[1][]{\ensuremath{\dot{\sv}}}
\newcommand{\dotsvd}[1][]{\ensuremath{\dot{\svd}}}
\newcommand{\dotsvp}[1][]{\ensuremath{\dot{\svp}}}

\newcommand{\deltasv}[1][]{\ensuremath{\Delta\sv}}
\newcommand{\deltasvd}[1][]{\ensuremath{\Delta\svd}}
\newcommand{\deltasvp}[1][]{\ensuremath{\Delta\svp}}

\newcommand{\y}[1][]{\ensuremath{\vctr y}}

\newcommand{\Burgers}[1]{\ensuremath{\vctr b^{#1}}}
\newcommand{\n}[1]{\ensuremath{\vctr n^{#1}}}

\newcommand{\dotgalpha}{\ensuremath{\dot{\gamma}^{\alpha}}}

\begin{document}
\let\WriteBookmarks\relax
\def\floatpagepagefraction{1}
\def\textpagefraction{.001}

\shorttitle{Phase-Field Modeling of Coupled Brittle-Ductile Fracture...}    

\shortauthors{S. Vakili et al.}  


\title [mode = title]{Phase-Field Modeling of Coupled Brittle-Ductile Fracture in Aluminum Alloys}


\let\printorcid\relax
\author[1]{\textcolor{black}{Samad} Vakili}[]

\cormark[1]


\ead{s.vakili@mpie.de}


\credit{Conceptualization, Data curation, Formal analysis, Investigation, Methodology, Software, Validation, Visualization, Writing-original draft, Writing-review and editing}

\affiliation[1]{organization={Microstructure Physics and Alloy Design, Max-Planck-Institut f\"ur Eisenforschung},
            addressline={Max-Planck-Stra\ss e 1}, 
            city={D\"usseldorf},
            postcode={40237}, 
            state={},
            country={Germany}}

\author[2,3]{\textcolor{black}{Pratheek} Shanthraj}[]

\credit{Methodology, Software, Writing-review and editing}

\affiliation[2]{organization={The Department of Materials, The University of Manchester},
            addressline={}, 
            city={Manchester},
            postcode={M13 9PL}, 
            state={},
            country={UK}}
\affiliation[3]{organization={Henry Royce Institute for Advanced Materials, The University of Manchester},
            addressline={}, 
            city={Manchester},
            postcode={M13 9PL}, 
            state={},
            country={UK}}

\author[1]{\textcolor{black}{Franz} Roters}[]

\credit{Resources, Software, Data curation, Supervision, Writing-review and editing}

\author[1]{\textcolor{black}{Jaber R.} Mianroodi}[]
\cormark[1]

\ead{j.mianroodi@mpie.de}

\credit{Methodology, Software, Validation, Writing-review and editing}

\author[1]{\textcolor{black}{Dierk Raabe}}[]

\credit{Funding acquisition, Project administration, Resources, Supervision, Writing-review and editing}
\cortext[1]{Corresponding Authors}

\begin{abstract}
Fracture in aluminum alloys with precipitates involves at least two mechanisms, namely, ductile fracture of the aluminum-rich matrix and brittle fracture of the precipitates. In this work, a coupled crystal plasticity - phase field model for mixed ductile-brittle failure modes is formulated and used to investigate the effect of precipitate morphology and size distribution on damage evolution in aluminum alloys. A thermodynamically consistent framework for elastic and plastic work dissipation in the fracture process zone is used to formulate the coupled constitutive behavior for brittle and ductile damage, respectively. Representative Volume Elements (RVE) with varying particle morphology and orientation were generated and their uni-axial loading was simulated to assess the damage resistance of the different model microstructures. For critical energy release rate ($G_c$) values of the aluminum matrix ranging from 4 to 8~Jm$^{-2}$ for single crystals and from 10 to 16~Jm$^{-2}$ for polycrystals, the model predicts a change in failure modes from particle debonding to cracking followed by ductile matrix failure. The change of particle failure mechanism as a result of increased critical energy release rate is observed for both, single and polycrystalline model microstructures. For the case of particle debonding, microstructures with circular or ellipsoidal particles (with the major axis perpendicular to the loading direction) show a higher ductility and fracture work compared to the other studied cases. In the case of particle cracking, microstructures with ellipsoidal particles aligned parallel to the loading axis show a higher ductility and fracture work among the investigated cases. The simulations qualitatively reproduce the experimentally observed particle failure mechanisms (cracking and debonding) for two different matrix alloy classes (commercially pure and 2xxx alloys) reinforced with ceramic particles. 
\end{abstract}


\maketitle

\section{Introduction}\label{sec:Intro}
The use of aluminum alloys in the transportation sector is projected to grow, due to the increased importance of light-weight materials and recyclability in the global agenda to reduce carbon dioxide emissions. However, the deployment of recycled aluminum alloys with higher impurity content and a tolerable fraction of undesired intermetallic particles is currently impeded by their poor formability and in-service ductility. Understanding the relationship between the microstructure of alloys and their damage tolerance \cite{Tasan2014,Wang2018,Bieler2009} is therefore critical to the development of aluminum alloys that are made from scrap \cite{Raabe2022}. The predominant failure mechanism in aluminum alloys is ductile fracture, with damage initiating around large precipitates due to strain localization and high stress triaxiality \cite{Lassance2007,Hannard2016,Shen2013}. This is exacerbated in recycled alloys, since the intermetallic particles, which are formed due to the presence of higher amounts of scrap-related impurities, are regions of strain localization. The localized strain and plastic deformation is detrimental for alloy performance in terms of ductility and damage tolerance. Experimentally investigating the effects of precipitates on the mechanical behavior of different aluminum alloys is demanding and time consuming. Employing computational methods for studying the fracture behavior of aluminum alloys can therefore be an essential additional approach which particularly allows to separate different effects on the structure-property relations, such as particle shape, decohesion, number density, size or dispersion.

Several studies have addressed the effects of precipitates and/or impurities on the fracture behavior of aluminum alloys using experiments. The ductile fracture of alloys with brittle particles is initiated by nucleation of micro-voids. In aluminum alloys, particle fracture or debonding from the matrix are as major mechanisms of damage initiation \cite{Argon1975,Lassance2007,Babout2004,Hannard2016}. Lassance et al. \cite{Lassance2007} have observed both, particle breakage and particle/matrix debonding in in-situ tensile tests of AA 6060 alloys. Similarly, Babout et al. \cite{Babout2004} have shown these two damage mechanisms for two aluminum alloys, commercially pure Al and Al2124 alloy, reinforced with ceramic particles. Effects of particle size and distribution on the fracture behavior of Al 6xxx alloys have been studied by Hannard et al. \cite{Hannard2016,Hannard2017}. They showed that the particle size distribution plays an important role for the ductility of such alloys \cite{Hannard2016}. Friction stir processing has been studied as a way to improve the fracture strain of Al 6056 alloys through homogenisation of the particle distribution, which yielded a shift of the size distribution towards smaller values \cite{Hannard2017}.

Several damage modeling studies have been published, dealing with aluminum composites and the effect of particulates on their fracture behavior \cite{Geni1998,Law2012,Babout2004,Segurado2003,Segurado2006,Ayyar2008}. Geni et al. \cite{Geni1998} have used a Finite Element Method (FEM)-Gurson model to study the influence of the size distribution of SiC particles on the damage behavior. They showed that a uniform size distribution of these particles leads to a higher fracture strain. The role of particle clustering on the mechanical properties of Metal Matrix Composites (MMCs) has been investigated by micro-mechanical modeling \cite{Segurado2003,Segurado2006} and it has been found that even a small perturbation of the particles' spatial homogeneity significantly increases their damage probability. Moreover, it has been shown that the damage preferentially initiates from particles existing close to each other. Similar results about the damage initiation and particle dispersion have been also obtained by others \cite{Ayyar2008,Mishnaevsky2004}. However, in case of nano-sized particles, the mechanical response of MMCs change dramatically \cite{Law2012}. Using discrete dislocation simulations, it was shown that the mechanical response of the nano-composite is significantly deteriorated by the damage of non-clustered and weakly clustered particles \cite{Law2012}.

For modeling ductile fracture, which is the main mechanism for the failure of aluminum alloys, the Gurson approach \cite{Gurson1977} and its extensions \cite{Tvergaard1984} have been applied to model the nucleation and growth of the micro-voids which leads to creation of micro-cracks or voids via coalescence. Formation of micro-voids or micro-cracks is driven by strain localization, due to presence of hard inclusions in the microstructure. Based on the characteristics of the inclusions and the associated microstructure, formation of either micro-cracks via brittle fracture of inclusions or micro-voids via debonding of the inclusions from the base microstructure initiate the damage process. 

Modeling approaches for ductile fracture fall into two categories, where cracks or voids are modeled either by discontinuous or by continuous methods. Discontinuous models include the FEM \cite{Ran2013}, Cohesive Zone formulations (CZ) \cite{Needleman1990}, and extended FEM (XFEM) \cite{Moes1999} methods. Continuous approaches include continuum damage models, damage gradient, and Phase Field (PF) models. 
In contrast to the discontinuous approaches where the crack is explicitly modeled as a discontinuity in the material, in continuous approaches it is modeled via gradual material degradation during the damaging process, for instance in terms of the damage parameter in the PF method.

During the last years, the PF method has emerged as a promising approach for modeling fracture \cite{Bourdin2000,Kuhn2010,Spatschek2011,Miehe2010,Borden2014,Ambati2015,Shanthraj2016,Shanthraj2017,Borden2012,Miehe2015,Aygun2021,REZAEI2021104253}, besides its widespread application for the simulation of phase transformation and microstructure evolution \cite{Boettinger2002,Chen2002,Steinbach2009,Tourret2022,Vakili2020,Vakili2021,Mianroodi2022}. 
In the PF brittle fracture models \cite{Bourdin2000,Miehe2010,Kuhn2010,Borden2014,Shanthraj2016,Shanthraj2017,Borden2012}, the discontinuity caused by the presence of a crack is represented by a non-conserved PF order parameter, which varies smoothly from a cracked surface ($\phi=0$) to the intact material ($\phi=1$), characterized by a specified length scale. The initiation and propagation of the crack is driven by the minimization of the total free energy, considering contributions from elastic strain release (gain in energy) and surface energy of the crack (loss in energy). The energetic description of the brittle fracture PF models reduces to the classical Griffith theory when the characteristic length scale approaches zero. 
When considering elasto-plastic materials, corresponding PF models have to also take into account the inelastic deformation of the material, which can be described in the framework of the finite strain approximation \cite{Shanthraj2016,Shanthraj2017,Borden2012,Miehe2015}.

While many studies exist on brittle fracture, only few works addressed the modeling of ductile fracture, for example \cite{Ambati2015,Miehe2016,Borden2016,Miehe2017,Aldakheel2018,Noll2020,Dittmann2018}. Only some of these formulations use the finite deformation approximation of the material, for instance \cite{Borden2016,Dittmann2018,Miehe2016,Miehe2017,Aldakheel2018}. The model proposed by Ambati \cite{Ambati2015} couples the damage degradation function with the accumulation of an effective plastic strain. For large strains, Miehe et al. \cite{Miehe2016} presented a PF model for ductile fracture, where the crack propagation is based on the accumulated effective elasto-plastic work. They assume that the hydrostatic pressure has negligible effect on material failure under plastic deformation conditions and that the gradual damage evolution reduces both, the material's total elastic stiffness (due to gradual loss of material coherency) and the material's strain hardening capability. In a similar approach, Borden et al. \cite{Borden2016} presented a ductile PF fracture model in which damage propagation reduces the size of the material's yield surface. Moreover, they addressed the stress triaxiality as a driving force for damage propagation. An approach to integrate the Gurson model in the PF modeling of ductile fracture was introduced by Aldakheel et al. \cite{Aldakheel2018}, where the criterion for damage initiation and propagation is based on the void volume fraction. In spite of the rich variety of existing ductile fracture models, there are yet no crystal plasticity ductile fracture models available for materials undergoing large deformations.

Therefore, in this paper a new PF model is formulated to simulate ductile fracture in conjunction with a crystal plasticity formulation. It is coupled with an existing brittle fracture PF model \cite{Shanthraj2016} to study damage evolution in aluminum alloys which contain stiff particles, such as precipitates, intermetallic compounds or non-metallic inclusions, that can intrude from scrap or processing. The ductile damage model is applied to the soft aluminum matrix, while the brittle damage model is used to simulate damage in the stiff particles. In the brittle damage model \cite{Shanthraj2016}, the initiation of cracks effectively degrades the materials stiffness such that the fully cracked region looses its cohesion and thus its stiffness entirely. In contrast, in the ductile model part cracking of the material effectively leads to yield degradation.

The paper is organized as follows: the theory is summarized in section~\ref{sec:theory}, the details of the simulations as well as the model parameters are introduced in section~\ref{sec:simulation setup}. The result of the simulations for aluminum alloys with Mg$_2$Si precipitates are discussed in section~\ref{sec:results and discussion}, followed by conclusions. 

\section{Theory}
\label{sec:theory}

\subsection{Kinematics}
We define a microstructural domain $\domain B_0 \subset \tnsrfour R^3$ with $\partial \domain B_0$ as the boundary.
The deformation under an applied load can be defined by a mapping function $\defmap(\vctr x):\vctr x \in \domain B_0 \to \vctr y \in \domain B$ from points \vctr x in the reference configuration $\domain B_0$ to points \vctr y in the deformed configuration \domain B. Here, we use the multiplicative decomposition of the deformation gradient, given by $\F = \partial\defmap/\partial\vctr x = \Grad \defmap$, into the elastic and plastic components, $\F = \Fe\Fp$. The elastic part of the deformation gradient \Fe\ is used to obtain the second \term{Piola--Kirchhoff} stress with $\sPK = \tnsrfour C\,(\transpose{\Fe}\Fe-\eyetwo)/2$, where \tnsrfour{C} is the elastic stiffness tensor. For a given plasticity model, the plastic velocity gradient $\Lp = \Fpdot\inverse{\Fp}$ is evolved using \sPK.

\subsection{Thermodynamics}
We confine the model formulations to isothermal and quasi-static processes with no external heat supply. The balance laws for linear momentum, angular momentum, internal energy, and total entropy are given by 
\begin{equation}
\bm{0}=\Divergence\,\fPK
\,,\quad
\F \transpose{\fPK} = \transpose{\F} \fPK
\,,\quad
\dot{\varepsilon}
=\fPK\cdot\dot{\F}
\,,\quad
\dot{\eta}
=\pi
\,,
\label{equ:BalMomEneEntDam}
\end{equation}
where, \fPK\ is the \term{Piola--Kirchhoff} stress, $\varepsilon$ is the internal energy density, $\eta$ represents the entropy density, and $\pi$ is the entropy production-rate density. 
Through the absolute temperature $\theta$, the entropy balance (Eq.~\ref{equ:BalMomEneEntDam}${}_{4}$) can be written in the corresponding dissipation balance form 
\begin{equation}
\theta\dot{\eta}=\delta,
\label{equ:BalDisDam}
\end{equation}
where $\delta:=\theta\pi$ represents the dissipation-rate density. 
Combination of Eqs.~\ref{equ:BalMomEneEntDam}${}_{3}$ and~\ref{equ:BalDisDam} then yields the reduced form 
\begin{equation}
\delta
= \fPK \cdot \Grad \dot{\defmap}
-\dot{\psi}
\label{equ:DenRatDisDam}
\end{equation}
of the dissipation-rate density \(\delta\), again at constant temperature, 
where $\psi:=\varepsilon-\theta\,\eta$ represents the free energy density. 

All free energy density models to be discussed below represent 
special cases of the general form 
\begin{equation}
\psi = \psi\left(\Grad \defmap,
\Fp,
\vctrgreek{\xi},
\varphi,
\Grad\varphi\right)
\label{equ:DenEneFreGen}
\end{equation}
in terms of the inelastic local deformation, $\Fp$, a set of local internal variables, $\vctrgreek{\xi}$, and the scalar damage order parameter \(\varphi\). 

Modeling the stress as purely energetic, we have 
\begin{equation}
\fPK = \partial_{\Grad \defmap }\psi
\label{equ:RelConDepEneDam}
\end{equation}
for the first \term{Piola--Kirchhoff} stress. In this case, 
\begin{equation}
\begin{array}{rcl}
\displaystyle
\int_{\domain B_0}
\delta \,
\,\inc{\vctr{x}}
&=&
\displaystyle
-\int_{\domain B_0}^{}
\left[
\partial_{\varphi}\psi
\ \dot{\varphi}
+\partial_{\Grad \varphi}\psi
\cdot\Grad \dot{\varphi}
+\partial_{\Fp}\psi
\cdot
\dot{\Fp}
+\partial_{\vctrgreek{\xi}}\psi
\cdot
\dot{\vctrgreek{\xi}}
\right]
\,\inc{\vctr{x}}
\\[3mm]
&=&
\displaystyle
-\int_{\domain B_0}^{}
\left[
\delta_{\varphi}\psi
\ \dot{\varphi}
+\partial_{\Fp}\psi
\cdot
\dot{\Fp}
+\partial_{\vctrgreek{\xi}}\psi
\cdot
\dot{\vctrgreek{\xi}}
\right]
\,\inc{\vctr{x}}
-\int_{\partial \domain B_0}
\partial_{\Grad \varphi}\psi
\cdot
\vctr{n}
\,\dot{\varphi}
\,\inc{\vctr{s}}
\,\\
\end{array}
\label{equ:RatDisDam}
\end{equation}
follows from Eq.~\ref{equ:DenEneFreGen} for the dissipation rate with 
respect to $\domain B_0$, where 
\begin{equation}
\delta_{x} f
=\partial_{x} f
- \Divergence \partial_{\Grad x} f
\label{equ:DerVar}
\end{equation}
represents the variational derivative. Assuming no-flux (Neumann) 
or constant-rate (Dirichlet) boundary conditions, \ie, 
\begin{equation}
\partial_{\Grad \varphi} \psi \cdot \vctr{n}
\vert_{\partial \domain B_0}
=0
\,,\quad
\dot{\varphi}
\vert_{\partial \domain B_0}
=0
\,,
\label{equ:ConBouFluNoConRatDam} 
\end{equation}
respectively, Eq.~\ref{equ:RatDisDam} reduces to the bulk form 
\begin{equation}
\delta
= - \delta_{\varphi} \psi
\ \dot{\varphi}
- \partial_{\Fp} \psi
\cdot
\dot{\Fp}
-\partial_{\vctrgreek{\xi}} \psi
\cdot
\dot{\vctrgreek{\xi}}
\label{equ:DenRatDisRedDam}
\end{equation}
for the dissipation rate and its density. 

To model kinetics in the context of Eq.~\ref{equ:DenRatDisRedDam}, 
attention is restricted here to constitutive relations based on the 
dissipation potential 
\begin{equation}
\chi = 
\chi(
\partial_{\Fp}\psi,
\partial_{\vctrgreek{\xi}}\psi,
\delta_{\varphi}\psi).
\label{equ:PotForDisDam}
\end{equation}
This potential determines the dependent constitutive quantities 
\begin{equation}
\dot{\varphi}
=-\delta_{\smash{\delta_{\varphi}\psi}}^{}\chi
\,,\quad
\dot{\Fp}
=-\partial_{\smash{\partial_{\Fp}\psi}}\chi
\,,\quad
\dot{\vctrgreek{\xi}}
=-\partial_{\smash{\partial_{\vctrgreek{\xi}}\psi}}\chi
\,,
\label{equ:RelEvoPotForDam}
\end{equation}
so that the residual form \(\delta\) via Eq.~\ref{equ:DenRatDisRedDam} becomes
\begin{equation}
\delta
=\delta_{\varphi}\psi
\ \delta_{\smash{\delta_{\varphi}\psi}}\chi
+\partial_{\smash{\Fp}}\psi
\cdot
\partial_{\smash{\partial_{\Fp}\psi}}\chi
+\partial_{\smash{\vctrgreek{\xi}}}\psi
\cdot
\partial_{\smash{\partial_{\vctrgreek{\xi}}\psi}}\chi
\label{equ:DenRatDisPotForDam}
\end{equation}
Assuming further that \(\chi\) is non-negative (\(\chi\geqslant 0\)) and convex 
in the thermodynamic driving forces, \ie, 
\begin{equation}
\delta_{\varphi}\psi
\ \delta_{\smash{\delta_{\varphi}\psi}}\chi
+\partial_{\Fp}\psi
\cdot
\partial_{\smash{\partial_{\Fp}}\psi}\chi
+\partial_{\vctrgreek{\xi}}\psi
\cdot
\partial_{\smash{\partial_{\vctrgreek{\xi}}\psi}}\chi
\geqslant 
\chi
\,,
\end{equation}
the resultant constitutive form expressed by Eq.~\ref{equ:DenRatDisPotForDam} fulfils the dissipation principle \(\delta\geqslant 0\).

\subsection{Constitutive modeling}
\label{sec: constitutive}

Using the general formalism outlined above, a constitutive law is now reduced to specifying two potentials: the free energy, $\psi$, and the dissipation potential, $\chi$, along with a suitable parameterization of the microstructure as a set of internal state variables. 
In the current constitutive model, the total free energy is separated into elastic, plastic and damage contributions.
\begin{equation}
\psi = \psi_\mathrm{E} + \psi_\mathrm{P} + \psi_\mathrm{D}
\end{equation}
These terms are explained in the context of the ductile fracture model in the following section

\subsection{Ductile fracture}

Ductile fracture occurs as a result of plastic deformation and its localization \cite{Pineau2007}. Therefore, the plastic work provides a major contribution to the total free energy. The elastic strain energy contribution is in most ductile materials smaller than the plastic work, owing to their usually moderate elastic modulus and the high local inelastic deformation that precedes damage initiation.
Similarly, the elastic contribution to the energy release rate in the fracture zone is significantly lower than that of the plastic energy. However, for the sake of completeness and for treating also materials with high elastic stiffness, we consider the elastic contribution in the ductile fracture here as well. 
The elastic free energy is given by
\begin{equation}
\psi_\mathrm{E} = \varphi^2 \tilde{\psi}_\mathrm{E},
\end{equation}
where $\tilde{\psi}_\mathrm{E}$ is the elastic strain energy of the intact (undamaged) material, which is given by $\tilde{\psi}_\mathrm{E}= \sPK\cdot\Ee/2$, with $\sPK = \tnsrfour C\,\Ee$ and $\Ee = (\transpose{\Fe}\Fe-\eyetwo)/2$.

We use in the present work a phenomenological crystal plasticity formulation for alloys with Face-Centered Cubic (FCC) crystal structure \cite{Hutchinson1976,Peirce1982}. In this model, the internal state variables, $\vctrgreek{\xi}$, associated with plasticity are parameterised in terms of the accumulated plastic slip $\gamma^\alpha$ on each of the crystallographic slip systems, which are indexed by $\alpha = 1,\ldots,N_{sl}$.
The thermodynamic driving forces conjugate to \Fp\ is given by
\begin{equation}
\partial_{\Fp} \psi= \partial_{\Fp} \psi_\mathrm{E} = - \sPK\, \invtranspose{\Fp}.
\end{equation}

The plastic deformation is described by activation of the slip systems when the resolved shear stress $\tau^\alpha$ exceeds the critical resolved shear stress $g^\alpha$. For a more detailed description of the model see \cite{Eisenlohr2013}. 

The driving force for ductile fracture is the degradation of the stored plastic energy, \ie\ the local loss in work hardening, given by
\begin{equation}
\psi_\mathrm{P} = \varphi^2 \sum_{\alpha}\int{g^\alpha \inc{\gamma^\alpha}},
\end{equation}
from the undamaged state, \ie\ $\varphi = 1$, to the fully damaged state, \ie\ $\varphi = 0$. The thermodynamic driving force conjugate to the internal variables $\gamma^\alpha$ is given by $\partial_{\gamma^\alpha} \psi = \varphi^2 g^\alpha$, \ie\ the slip resistance on slip system $\alpha$.

Using a power law dissipation function for the set of yield surfaces $-\partial_{\Fp} \psi \transpose{\Fp} \cdot (\Burgers{\alpha}\otimes\n{\alpha}) - \partial_{\gamma^\alpha} \psi \le 0$, the plastic constitutive equations are given by
\begin{equation}
\Fpdot\inverse{\Fp} = \Lp = \sum_{\alpha} \dotgalpha \, \Burgers{\alpha}\otimes\n{\alpha}, \quad\text{and}\quad \dot\gamma^\alpha = \dot\gamma_0\abs{\frac{\tau^\alpha}{\varphi^2 g^\alpha}}^n \sign{\tau^\alpha},
\end{equation}
where $\Burgers{\alpha}$ and $\n{\alpha}$ are unit vectors along the slip direction and slip plane normal, respectively, $\tau^\alpha = \sPK \cdot (\Burgers{\alpha}\otimes\n{\alpha})$ is the resolved shear stress, $\dot \gamma_0$ is the reference shear rate, and $n$ the stress exponent.

The dependence of the plastic free energy, $\psi_\mathrm{P}$, on $\gamma^\alpha$ is determined by the form of the hardening law. 
The slip resistances on each slip system, $g^\alpha$, evolve asymptotically towards $g_\infty$ with shear $\gamma^\beta$ ($\beta = 1,\ldots,12$) according to the relationship
\begin{align}
	\label{eq: hardening pheno}
	\dot g^\alpha &= \dot\gamma^\beta \, h_0 \left| 1 - g^\beta/g_\infty \right|^a \sign{1 - g^\beta/g_\infty} h_{\alpha\beta},
\end{align}
with the parameters $h_0$ and $a$.
The interaction between different slip systems is captured by the hardening matrix $h_{\alpha\beta}$.

In the current ductile damage model, the damage contribution to the free energy is given by
\begin{equation}
\psi_\mathrm{D} = \frac{G_c}{l_{0}} (1 - \varphi) + \frac{1}{2}\,l_{0}\,G_{c}\,|\nabla\varphi|^{2},
\label{equ:damageE}
\end{equation}
where $G_c$ is the critical energy release rate in the ductile fracture model. The thermodynamic driving forces conjugate to the damage phase field can be obtained as
\begin{equation}
\delta_{\varphi}\psi = \partial_{\varphi}\psi_\mathrm{E}+\partial_{\varphi}\psi_\mathrm{P} + \delta_{\varphi}\psi_\mathrm{D} = 2\varphi \Big( \tilde{\psi}_\mathrm{E} + \sum_{\alpha}\int{g^\alpha \inc{\gamma^\alpha}} \Big) - \frac{G_c}{l_{0}} - \Divergence \,l_{0}\,G_{c}\Grad \varphi.
\end{equation}
The constitutive description is completed by specifying the following convex, non-negative form of the dissipation potential, 
\begin{equation}
\chi_{\mathrm{D}} = \frac{1}{2} M \left( \delta_{\varphi}\psi \right)^2,
\label{equ:dissED}
\end{equation}
where $M$ is a mobility parameter, which quantifies at which rate the decohesion zone is opening up, quantified by the phase-field variable, until its value reaches zero, translating to a fully developed crack.
Equation.~\ref{equ:dissED} results in the following field equation for the evolution of $\varphi$
\begin{equation}
\dot{\varphi} = -M \left[2\varphi \Big( \tilde{\psi}_\mathrm{E} + \sum_{\alpha}\int{g^\alpha \inc{\gamma^\alpha}}\Big) - \frac{G_c}{l_{0}} - \Divergence \,l_{0}\,G_{c}\Grad \varphi\right],
\label{equ:damagePF}
\end{equation}
%

\subsection{Brittle fracture}

The brittle fracture model used in this work follows the PF damage model described in \cite{Shanthraj2016}. The approach is based on the extension of the brittle fracture theory of Griffith, which assumes that the fracture process is driven by the elastic energy release rate in the process zone. The elastic free energy contribution is given by $\psi_\mathrm{E} = \varphi^2\tilde{\psi}_\mathrm{E}$, where the elastic energy degrades from the undamaged, \ie\ $\varphi = 1$ ($\psi_\mathrm{E} = \tilde{\psi}_\mathrm{E}$), to the fully damaged state, \ie\ $\varphi = 0$ ($\psi_\mathrm{E} \approx0$), state. The $\tilde{\psi}_\mathrm{E} =\sPK \cdot\Ee/2$ is the elastic strain energy of the undamaged material. 

The damage evolution equation for the brittle fracture model is given by

\begin{equation}
	\dot{\varphi} = - M \left[ \varphi \tilde{\psi}_\mathrm{E} - \frac{G_c}{l_{0}} - \Divergence \,l_{0}\,G_{c}\Grad \varphi \right],
	\label{eq:brittlefracture}
\end{equation}
where $M$ and $l_0$ are the damage mobility and the damage length scale, respectively. It is noteworthy that the material's effective stiffness $\mathbf{C}$ is gradually reduced by the progressing damage, as quantified by the change in the value of the damage parameter, \ie\ $\mathbf{C} = \phi^2 \tilde{\mathbf{C}}$. This means that the damage parameter also quantifies the material's loss in cohesion, which implies that for a fully damaged material, \ie\ $\phi=0$, the stiffness approaches zero. See \cite{Shanthraj2016} for more details about the model. 

\section{Simulation setup}
\label{sec:simulation setup}
The simulations are performed on a model aluminum alloy system containing Mg$_2$Si particles. Regarding the role of particle size, one should note, that it is particularly the larger ones that are more prone to initiate damage. In Al 6xxx series alloys, the average size of the particles that tend to initiate damage, including Mg$_2$Si, is around 1-20$\mu$m \cite{Hannard2016,Petit2019}. Therefore, in this work, we only consider those particles that are larger than a few microns. Sub-micron sized particles are not included explicitly. However, their effect on the material's strength is included through the constitutive description of the mechanical response of the aluminum matrix, see the model parameters given in Table~\ref{key}. For the case of alloys in which sub-micron sized particles are the largest ones that are present in the material, the model parameters, including the length scale parameter $l_0$ in \eqref{equ:damagePF} and \eqref{eq:brittlefracture}, can be correspondingly adjusted. In the following, the detail of the simulation setup is explained.

\begin{figure}
	\centering
	\includegraphics[width=0.32\linewidth]{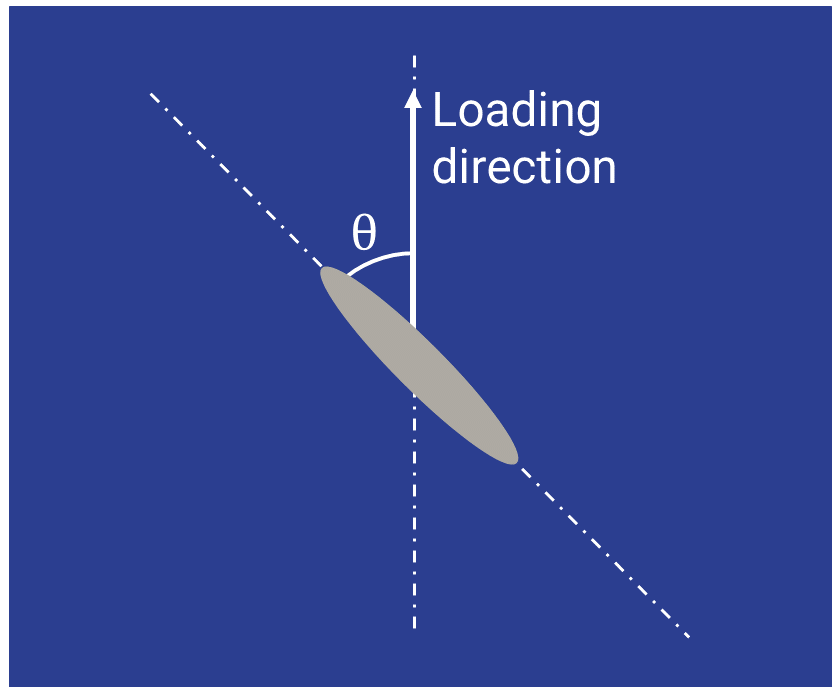}
	\caption{Schematic representation of ellipsoidal particles and the angle ($\theta$) between their longitudinal axis and the externally imposed loading direction. $\theta =0$, $45$, and $90^\circ$ are referred to as Parallel, $45^\circ$ Inclined, and Perpendicular, respectively. The loading direction, as shown here, is identical for all the simulations in this work.}
	\label{fig:Schematic}
\end{figure}

A set of simulations is performed on a single crystal Aluminum model alloy with randomly distributed Mg$_2$Si particles in space. The simulations are based on the ductile fracture model for the aluminum matrix and the brittle fracture model for the Mg$_2$Si particles, as described in section~\ref{sec:theory}. To investigate the role of particle shape and orientation on fracture behavior, simulations are performed for both, circular and ellipsoidal particles. For the latter case, three different scenarios for ellipsoidal particles are investigated, with their major axes aligned either parallel, 45$^\circ$ inclined, or perpendicular to the loading direction, respectively, i.e. forming angles of 0, 45, and 90$^\circ$ to the loading direction. A schematic picture of an ellipsoidal particle and the angle between the loading direction and the long axis of the particle is shown in Fig.~\ref{fig:Schematic}.
It should be noted that throughout this paper the uni-axial loading direction is the same as shown in Fig.~\ref{fig:Schematic}. Unless otherwise stated, the system size for all simulations is $200\times 160\times1~\mu$m which is discretized into voxels of size $1\times1\times1~\mu$m. Three random seeds are considered for generating three spatially random particle distributions in each case of circular, parallel, 45$^\circ$ inclined, and perpendicular ellipsoidal particles. Three different seeds for each geometry are considered to investigate dependency of the results on the particles' spatial distribution. It needs to be mentioned that the clustering of particles is excluded by assuring a minimum spacing of 20~$\mu m$ between neighboring particles. Furthermore, the volume fraction of particles is approximately identical in all simulations for the single crystal.

Similar simulations are designed for the polycrystalline aluminum alloys consisting of 10 grains with random orientations. Similar to the case of the single crystal simulations, several Mg$_2$Si particles with a minimum inter-particle spacing of 20$~\mu$m are placed inside the microstructure in a way that no particle is on the grain boundaries. 
Analogous to the single crystal simulations, in the polycrystal case four particle configurations are considered, namely, with circular, and ellipsoidal particles aligned parallel, 45$^\circ$ inclined, and perpendicular to the loading direction. However, here only one random spatial distribution of the particles is considered in each of the geometrical cases. 

In all the simulations (either the single crystal or the polycrystal), the critical energy release rate for the Mg$_2$Si particles is set to $G_c^{\textrm{Mg$_2$Si}}=16$~Jm$^{-2}$. Different values for the matrix critical energy release rate are used for each set of simulations, to explore its effect on the fracture behavior of aluminum alloys. For the single crystal simulations, $G_c^{\textrm{Al-M}}=4 $ and $8$~Jm$^{-2}$, and for the polycrystal simulations, $G_c^{\textrm{Al-M}}=10 $ and $16$~Jm$^{-2}$ are employed. Here, the matrix critical energy release rate corresponds to that of the aluminum alloy excluding the micro-scale Mg$_2$Si precipitates. This value can be different even for the same alloy having various sub-micron sized precipitates (for example $\beta^{\prime\prime}$, $\beta^\prime$, or even Guinier-Preston (GP) zones) with various morphologies or characteristics (for instance chemical bonding, or coherency in case of GP zones) \cite{Chen2006}. Through all the simulations, the sample is loaded in one direction (uniaxial tensile test). The aluminum matrix deforms elasto-plastically, while the particles (Mg$_2$Si) deform purely elastically. The elastic properties of both, the Mg$_2$Si \cite{Jain2013} particles and aluminum \cite{Ma2018} matrix are listed in Table~\ref{key}. For the aluminum matrix, with its FCC crystal structure, there are 12 slip system given by $\{111\}\langle110\rangle$. The constitutive parameters for the phenomenological crystal plasticity model for the aluminum are taken from reference \cite{Eisenlohr2013} and are summarized in Table~\ref{key}.

\begin{table}
	\centering	
	\caption{Material parameters of the aluminum (elastic constants from \cite{Thomas1968} and plastic parameters \cite{Eisenlohr2013}) and Mg$_2$Si particles \cite{Jain2013}.}
	\begin{tabular}{cccccccccccc}
		\hline
		& $\mathbf{C}_{11}$ & $\mathbf{C}_{12}$ & $\mathbf{C}_{44}$ & $\dot{\gamma}$ & $\xi_0$ & $\xi_\infty$ & $h_0$ & a & n & $h_{\alpha \beta}$ & $h_{\alpha \beta}$ \\
		& \footnotesize (GPa) & \footnotesize (GPa) & \footnotesize (GPa) & \footnotesize $(s^{-1})$ & \footnotesize (MPa) & \footnotesize (MPa) & \footnotesize (MPa) & & & & \\
		\hline
		\textbf{Aluminum} & 106.75 & 60.41 & 28.34 & 0.001 & 40 & 210 & 20 & 2.25 & 20.0 & 1.0 & 1.4 \\
		\textbf{Mg$_2$Si} & 113.09 & 23.0 & 44.0 & - & - & - & - & - & - & - & - \\
	\end{tabular}
	\label{key}
\end{table}

\section{Results and discussion}
\label{sec:results and discussion}
\subsection{Single crystal}
\label{sec:single crystal}
\begin{figure}
	\centering
	\includegraphics[width=0.4\linewidth]{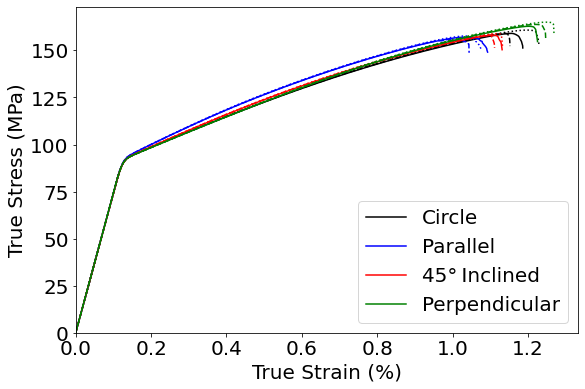}
	\includegraphics[width=0.4\linewidth]{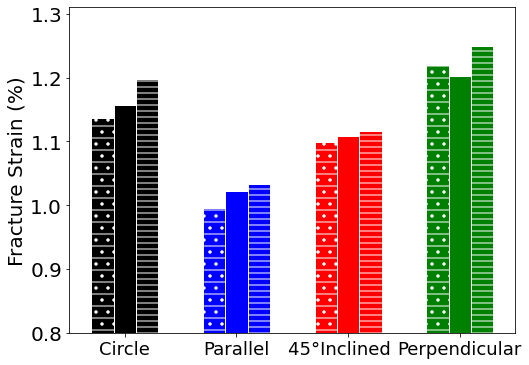}\\
	(a)\hspace*{0.4\linewidth}(b)\\
	\includegraphics[width=0.4\linewidth]{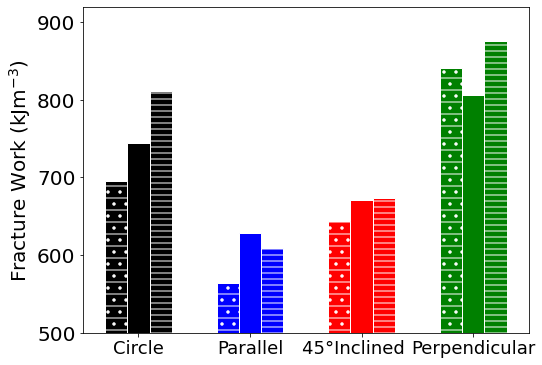}
	\includegraphics[width=0.4\linewidth]{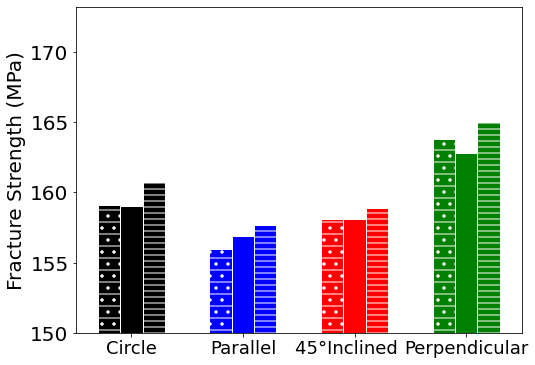}\\
	(c)\hspace*{0.4\linewidth}(d)
	\caption{(a) The stress-strain curves along the loading direction for the system of spatially random Mg$_2$Si particles inside the aluminum single crystal, where $G_c^{\textrm{Al-M}}=4$~Jm$^{-2}$ and $G_c^{\textrm{Mg$_2$Si}}=16$~Jm$^{-2}$. The colors correspond to the particles' shape for the circular one and their orientation about the loading direction for the ellipsoidal particles, respectively. The solid, dashed, and point lines in (a) represent three different randomly created spatial distributions of the particles. This is to show how the spatially random distribution of particles affects the result. (b) The fracture strain, (c) fracture work, and (d) fracture strength extracted from stress-strain curves. The different lines in the stress-strain curves are represented by different bars (solid, dashed, and pointed) in b-d.} 
	\label{fig:Circle_to_ellipse_StressStrainGc4}
\end{figure}

\begin{figure}
	\centering
	\includegraphics[width=0.4\linewidth]{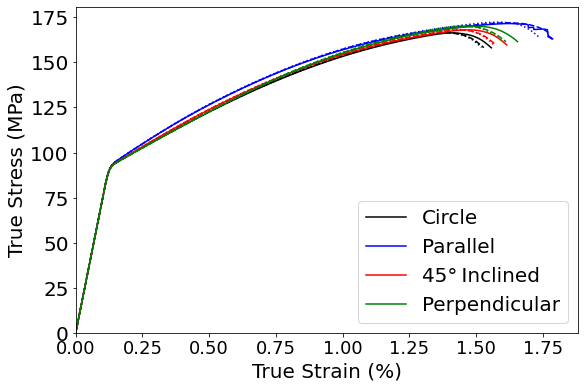}
	\includegraphics[width=0.4\linewidth]{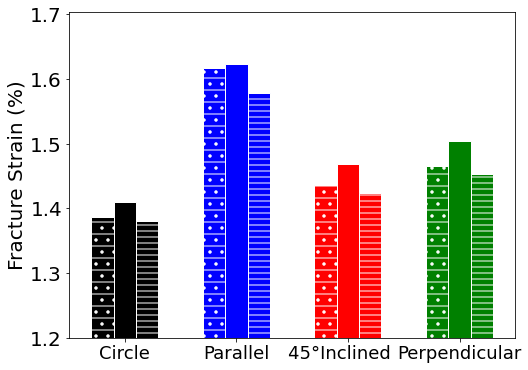}\\
	(a)\hspace*{0.4\linewidth}(b)\\
	\includegraphics[width=0.4\linewidth]{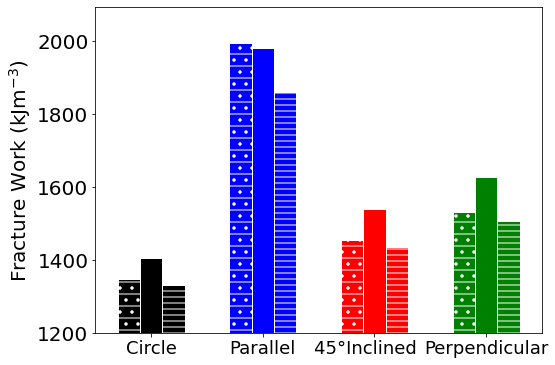}
	\includegraphics[width=0.4\linewidth]{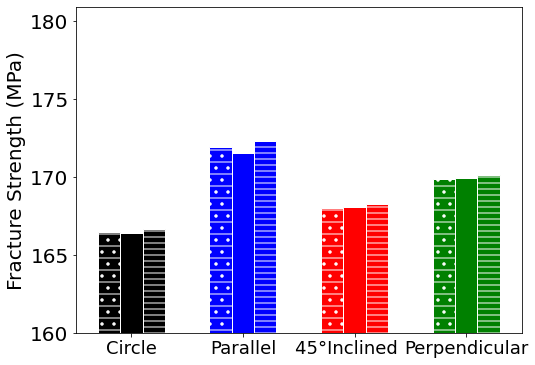}\\
	(c)\hspace*{0.4\linewidth}(d)
	\caption{ (a) The result of uniaxial tensile loading for the single crystal aluminum including different particle configurations (circular, and ellipsoidal particles aligned parallel, 45$^\circ$ inclined, and perpendicular to the loading direction), and (b-d) the extracted data similar to Fig.~\ref{fig:Circle_to_ellipse_StressStrainGc4} for higher value of aluminum critical energy release rate, $G_c^{\textrm{Al-M}}=8$~Jm$^{-2}$ and $G_c^{\textrm{Mg$_2$Si}}=16$~Jm$^{-2}$. The colors correspond to circular, and ellipsoidal particles aligned parallel, 45$^\circ$ inclined, and perpendicular to the loading direction. The solid, dashed, and point lines represent three different spatial distribution of the particles for each configuration. (a-d) Fracture strain, work, and strength which are extracted from stress-strain curves. The solid, dash, and point bars represent the three spatially random distribution of the particles for each configuration.}
	\label{fig:Circle_to_ellipse_StressStrainGc8}
\end{figure}

The uni-axial stress-strain response of the single crystal samples with four different particle configurations are depicted in Figs.~\ref{fig:Circle_to_ellipse_StressStrainGc4} and~\ref{fig:Circle_to_ellipse_StressStrainGc8} for $G_c^{\textrm{Al-M}}=4$ and $8$~Jm$^{-2}$, respectively, (see Section~\ref{sec:simulation setup}). Here, Al-M stands for the Aluminum Matrix. For the stress-strain curves in Figs.~\ref{fig:Circle_to_ellipse_StressStrainGc4}a and~\ref{fig:Circle_to_ellipse_StressStrainGc8}a, the colored lines represent either the shape (for the circular particles) or the orientation of particles about the loading direction (for ellipsoidal ones), respectively. For each figure, the extracted data for fracture strain, fracture work, and the fracture strength are shown in (b), (c), and (d), respectively. The solid, dashed, and pointed lines (Figs.~\ref{fig:Circle_to_ellipse_StressStrainGc4}c-d and Figs.~\ref{fig:Circle_to_ellipse_StressStrainGc8}c-d) correspond to three different spatially random distributions of the particles. Using the lower value of $G_c^{\textrm{Al-M}}=4$~Jm$^{-2}$, the alloys with ellipsoidal particles aligned parallel to the loading direction show the lowest ductility (fracture strain), fracture work, fracture stress, and fracture strength, see Fig.~\ref{fig:Circle_to_ellipse_StressStrainGc4}b-d, respectively. In contrast, for the alloys with higher matrix critical energy release rate $G_c^{\textrm{Al-M}}=8$~Jm$^{-2}$ (Fig.~\ref{fig:Circle_to_ellipse_StressStrainGc8}), the parallel particles aligned to the loading direction show superior mechanical behavior, higher fracture strain, fracture work, and higher fracture strength. It is noteworthy, that the alloys with parallel particles aligned to the loading direction show slightly higher strain hardening than the other particle configurations (circular, and ellipsoidal rather than those that are parallel to the loading direction) for both $G_c^{\textrm{Al-M}}=4$ and $8$~Jm$^{-2}$ (Figs.~\ref{fig:Circle_to_ellipse_StressStrainGc4}a and ~\ref{fig:Circle_to_ellipse_StressStrainGc8}a), and still they show the highest ductility, fracture work, and fracture strength for $G_c^{\textrm{Al-M}}=8$~Jm$^{-2}$ (Figs.~\ref{fig:Circle_to_ellipse_StressStrainGc8}b-d). This is in contrast with the fact that generally the ductility decreases as the strain hardening increases.

\begin{figure}
	\centering
	\includegraphics[width=0.4\linewidth]{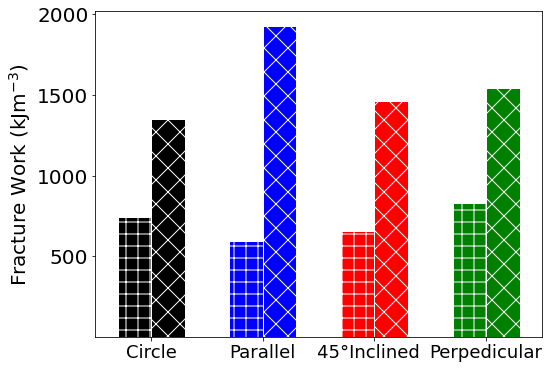}
	\caption{ The average fracture work for different particle configurations by using two different critical energy release rates for aluminum matrix $G_c^{\textrm{Al-M}}=4$ (shown with "+" bars) and 8~Jm$^{-2}$ (shown with "x" bars); $G_c^{\textrm{Mg$_2$Si}}=16$~Jm$^{-2}$. }
	\label{fig:Average_fracture_work}
\end{figure}

The comparison of the results for the two sets of single crystal simulations with two different values of $G_c^{Al-M}$ (Figs.~\ref{fig:Circle_to_ellipse_StressStrainGc4} and~\ref{fig:Circle_to_ellipse_StressStrainGc8}) reveals that the fracture strain and fracture work both increase when using a higher value of $G_c^{Al-M}=8$~Jm$^{-2}$, compared to $G_c^{Al-M}=4$~Jm$^{-2}$. The critical energy release rate, $G_c$ in Eqs.~\ref{equ:damagePF} (ductile fracture) and~\ref{eq:brittlefracture} (brittle fracture), determine the initiation of the damage. One might therefore expect that by doubling $G_c^\textrm{Al-M}$, the fracture work will likewise double since the critical energy release rate of particles $G_c^\textrm{Mg$_2$Si}$ is kept constant. In order to asses this estimate, the average values of the fracture work for different particle configurations are plotted in Fig.~\ref{fig:Average_fracture_work}, where the bars with "+" and "x" correspond to $G_c^\textrm{Al-M}=4$ and $8$~Jm$^{-2}$, respectively. The average fracture work approximately doubles with the $G_c^\textrm{Al-M}$ doubling, except for the case of parallel particles, where it almost triples.

\begin{figure}
	\centering
	\includegraphics[width=0.5\linewidth]{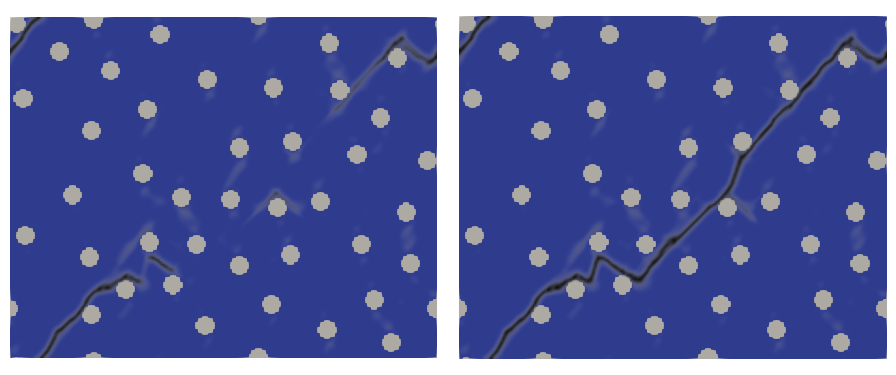}\\
	(a)\\
	\includegraphics[width=0.5\linewidth]{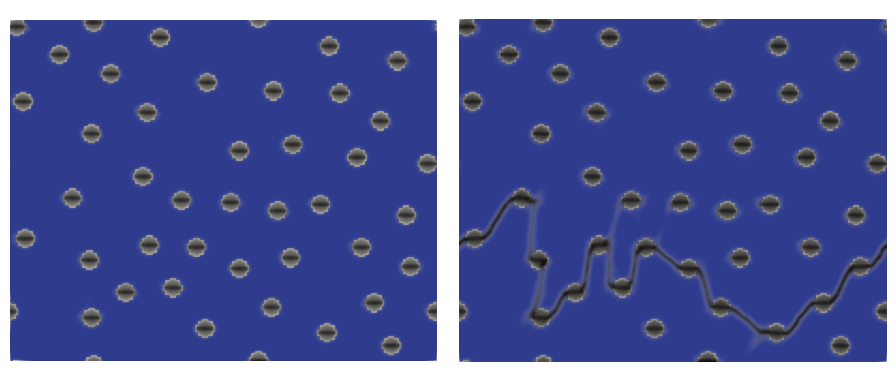}\\
	(b)
	\caption{Crack initiation and propagation (from left to right) for the same single crystal sample with two different critical energy release rates for the aluminum matrix, (a) $G_c^{\textrm{Al-M}}=4$ and (b) $8$~Jm$^{-2}$. The particle critical energy release rate is given by $G_c^{\textrm{Mg$_2$Si}}=16$~Jm$^{-2}$. The aluminum matrix is shown in blue and the particles are shown in grey. The cracked (damaged) region is represented by black color.}
	\label{fig:CrackPath_cricle_ellipse}
\end{figure}

\begin{figure}
	\centering
	\includegraphics[width=0.9\linewidth]{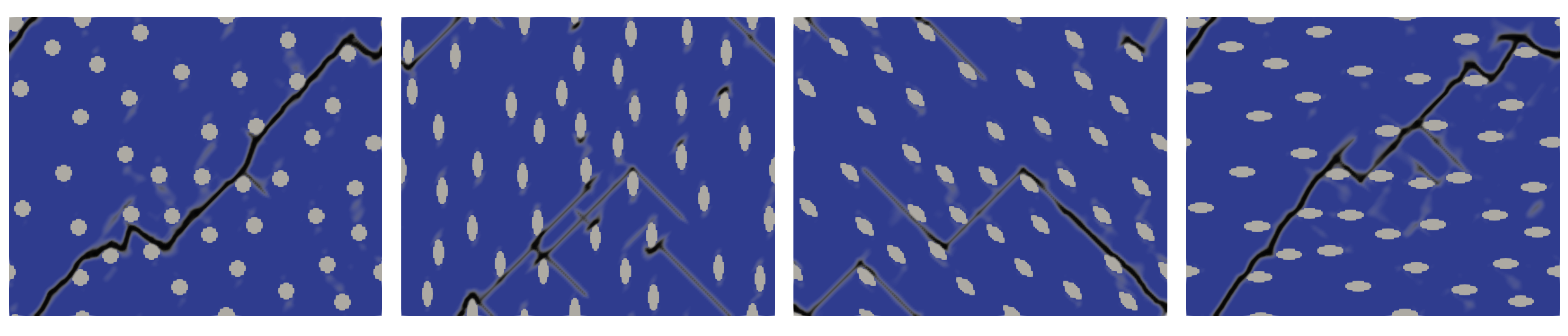}\\
	(a)\hspace*{0.21\linewidth}(b)\hspace*{0.21\linewidth}(c)\hspace*{0.21\linewidth}(d)\\
	\includegraphics[width=0.9\linewidth]{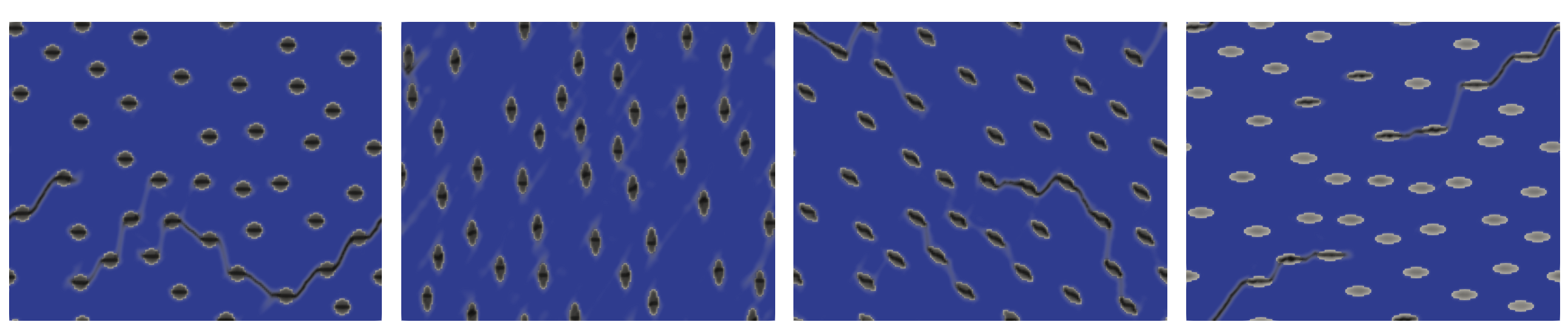}\\
	(e)\hspace*{0.21\linewidth}(f)\hspace*{0.21\linewidth}(g)\hspace*{0.21\linewidth}(h)
	)
	\caption{Crack propagation path for the samples with circular and ellipsoidal particles aligned parallel, 45$^\circ$ inclined, and perpendicular to the loading direction simulated by using $G_c^{\textrm{Al-M}}=4$~Jm$^{-2}$ (a-d) and $8$~Jm$^{-2}$ (e-h). For the lower value of $G_c^{\textrm{Al-M}}$ (a-d) the crack propagates along the particle/aluminum interface. In contrast, for higher $G_c^{\textrm{Al-M}}$ (e-h) cracking of the particle is the damage initiation cause and the crack propagates through the aluminum matrix to connect the particles cracks.}
	\label{fig:Circle_to_ellipseGc48}
\end{figure}

The variations in the fracture work can be explained by noting the microstructure and the damage mechanism. For the single crystal aluminum with circular particles, the crack evolution in Fig.~\ref{fig:CrackPath_cricle_ellipse} shows that the damage initiates from the Mg$_2$Si/aluminum matrix interface and propagates into the matrix for the case of $G_c^{Al-M}=4$~Jm$^{-2}$, Fig.~\ref{fig:CrackPath_cricle_ellipse}a, while cracking of Mg$_2$Si particles is the main mechanism for damage initiation in the alloys with $G_c^{Al}=8$~Jm$^{-2}$, see Fig.~\ref{fig:CrackPath_cricle_ellipse}b. For the former case, as shown in Fig.~\ref{fig:CrackPath_cricle_ellipse}a, besides damage initiation from the particle/aluminum interface, the crack tends to propagate along these interfaces until it sweeps through the sample. On the other hand, for $G_c^{Al}=8$~Jm$^{-2}$, crack propagation occurs by merging of the originally isolated cracks from the individually damaged particles, Fig.~\ref{fig:CrackPath_cricle_ellipse}b. The debonding of particle/matrix and cracking of the particles for the two kinds of aluminum alloys, with different critical energy release rates, are also observed for the rest of samples with different particle topology (ellipsoidal particles aligned parallel, 45$^\circ$ inclined, and perpendicular to the loading direction), see Fig.~\ref{fig:Circle_to_ellipseGc48}. The different types of damage mechanisms found for the two sets of simulations is the main reason for the observed variations in the fracture work ratios for the different particle configurations.

\clearpage
\subsection{Single crystal: Aluminum crystal orientation}

\begin{figure}
	\centering
	\includegraphics[width=0.45\linewidth]{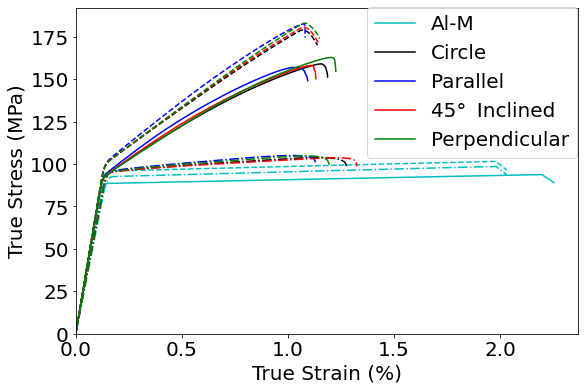}
	\includegraphics[width=0.45\linewidth]{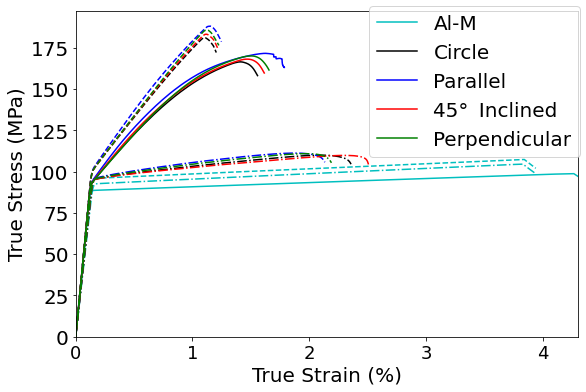}\\
	(a)\hspace*{0.45\linewidth}(b)
	\caption{ The effect of aluminum crystal orientation on the stress-strain response of the samples with different particle configurations, using two critical energy release rate for aluminum matrix (a) $G_c^{\textrm{Al-M}}=4$ and (b) 8~Jm$^{-2}$. $G_c^{\textrm{Mg$_2$Si}}=16$~Jm$^{-2}$. The solid lines correspond to the result of simulations for a given random crystal orientation. For point-dashed and dashed lines, the loading direction is parallel to $\langle010\rangle$ and $\langle110\rangle$ direction in the aluminum structure, respectively. The result of aluminum matrix without any particle (shown as Al-M) is also depicted for comparison purposes.}
	\label{fig:Crystal_orientation}
\end{figure}

The effect of the aluminum single crystal orientation on the damage behavior of the samples with different particle configuration is shown in Fig.~\ref{fig:Crystal_orientation} and compared with that of the aluminum matrix without any particles. 
The results of simulations with $G_c^{\textrm{Al-M}}=4$ and $8$~Jm$^{-2}$ are shown, respectively, in Figs.~\ref{fig:Crystal_orientation}a and~b. For both values of $G_c$, the curves depicted by the solid lines are the results that have already been discussed in Section~\ref{sec:single crystal}. Since the crystal orientation of theses simulations has been randomly selected with respect to the sample coordinates, we refer to them as 'random' in the following. This randomly chosen crystal orientation was used for all the samples with different particle configurations. The aim of showing these results is to compare them with the result obtained for the two other crystal orientations, namely, the cube and the rotated-cube orientation, which are depicted in terms of point-dashed lines and dashed lines, respectively in Fig.~\ref{fig:Crystal_orientation}. For the cube orientation, the crystal lattice coordinates coincide with the sample coordinates, while for the rotated cube, the crystal lattice is rotated 45$^\circ$ about the axis $a_3$ for the lattice represented by the three coordinates $a_1$, $a_2$, and $a_3$. Therefore, for the cube orientation, the uniaxial loading direction is parallel to the aluminum crystal orientation of $\langle010\rangle$, and for the rotated-cube crystal it is parallel to $\langle110\rangle$. Considering the crystal orientation (distinguished by line type) and particle topology (distinguished by color), the stress-strain response of the samples varies significantly with the crystal orientation rather than with the particles' shape/orientation, see Figs.~\ref{fig:Crystal_orientation}a and~b. In other words, for a single crystal, the effect of the crystal orientation of the host matrix on the mechanical response is more pronounced than that of the particles' configuration (shape/orientation). Regardless of the shape/orientation, the presence of particles by itself is the main reason for such a variety on the mechanical response observed for different aluminum crystal orientations, as seen in Figs.~\ref{fig:Crystal_orientation}a and~b. To explain this, for each crystal orientation, a simulation is performed for a matrix without any particle. The results are shown by turquoise color in Figs.~\ref{fig:Crystal_orientation}a and~b, where solid, point-dashed, and dashed lines correspond to the random, cube, and rotated cube orientation, respectively. As shown in Figs.~\ref{fig:Crystal_orientation}a and~b, the crystal orientation does not significantly change the stress-strain response of the samples without particles compared to those samples with particles. It is noteworthy, that for the cube orientation (loading along $\langle010\rangle$), the strain hardening of samples with different particle configurations (circular, and ellipsoidal aligned parallel, 45$^\circ$ inclined, and perpendicular to the loading direction) is similar to the samples without particles, but they show lower ductility. Although these simulations have been performed for a single crystalline host matrix, the results are also relevant for poly-crystalline materials. 
 
\begin{figure}
	\centering
	\includegraphics[width=0.45\linewidth]{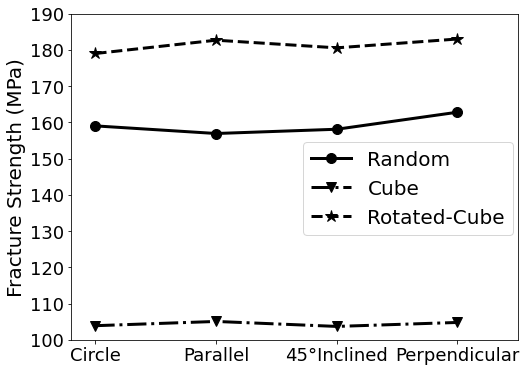}
	\includegraphics[width=0.45\linewidth]{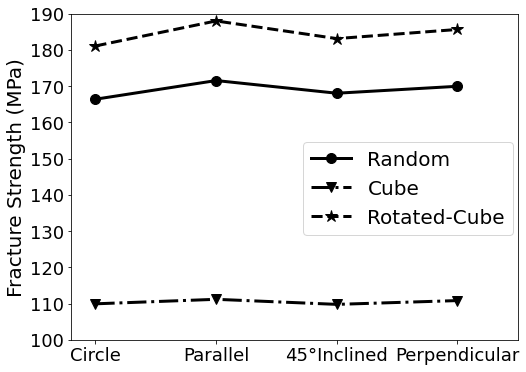}\\
	(a)\hspace*{0.45\linewidth}(b)\\
	\includegraphics[width=0.45\linewidth]{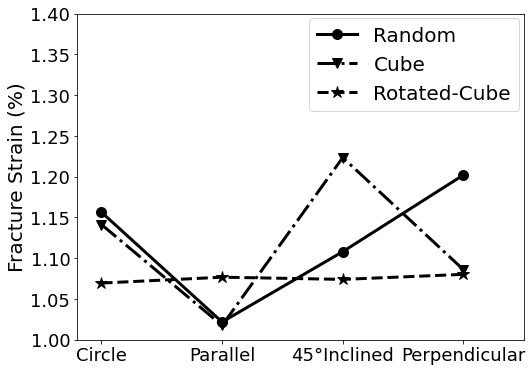}
	\includegraphics[width=0.45\linewidth]{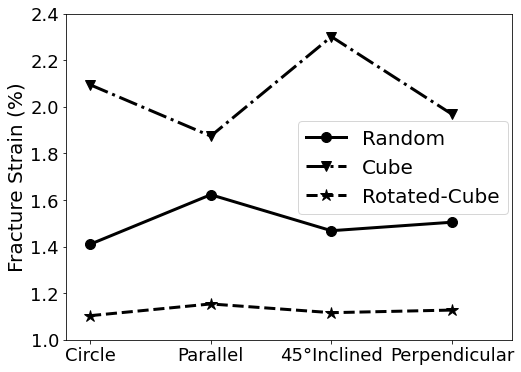}\\
	(c)\hspace*{0.45\linewidth}(d)\\
	\includegraphics[width=0.45\linewidth]{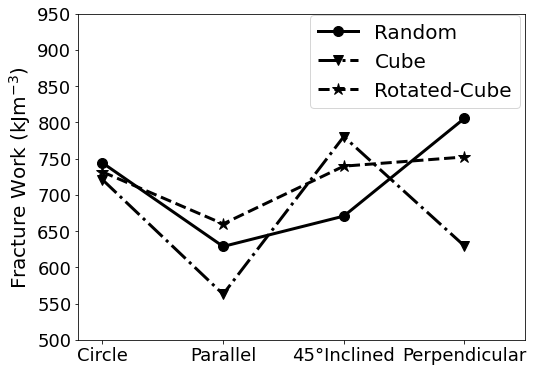}
	\includegraphics[width=0.45\linewidth]{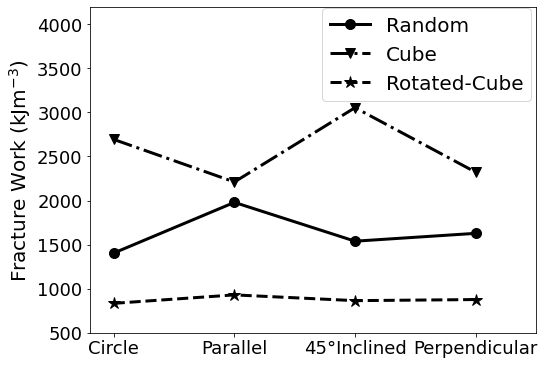}\\
	(e)\hspace*{0.45\linewidth}(f)
	\caption{ The extracted data (strength, fracture strain, and fracture work) from Fig.~\ref{fig:Crystal_orientation} for the simulations using using $G_c^{\textrm{Al-M}}=4$ (a, c, e) and using $8$~Jm$^{-2}$ (b, d, f). The first row show the results for maximum strength (a-b), while the second (c-d) and third (e-f) row show the fracture strain and fracture work, respectively. The solid lines correspond to the result of the simulations (from Section~\ref{sec:single crystal}) for a so-called random crystal orientation. The point-dashed and dashed lines show the result for the cube and rotated-cube orientations, respectively, where the loading direction is parallel to $\langle010\rangle$ and $\langle110\rangle$.}
	\label{fig:Crystal_orientation_extracted}
\end{figure}

To compare the stress-strain curves in Figs.~\ref{fig:Crystal_orientation}, the extracted data for maximum strength, fracture strain, and fracture work are plotted in Fig.~\ref{fig:Crystal_orientation_extracted}. The figures in the left (Fig.~\ref{fig:Crystal_orientation_extracted}a, c, and~e) are the result for $G_c^{\textrm{Al-M}}=4$~Jm$^{-2}$, and the ones on the right (Fig.~\ref{fig:Crystal_orientation_extracted}b, d, and~f) are for $G_c^{\textrm{Al-M}}=8$~Jm$^{-2}$. The maximum strength, fracture strain, and fracture work are shown, respectively, in the first (Fig.~\ref{fig:Crystal_orientation_extracted}a-b), second (Fig.~\ref{fig:Crystal_orientation_extracted}c-d), and third (Fig.~\ref{fig:Crystal_orientation_extracted}e-f) row. In all figures, the x-axis corresponds to different particle configurations, while different line styles show the aluminum matrix crystal orientation, solid line for random, point-dashed line for cube, and dashed line for rotated-cube orientation. The results referred to as 'random' have already been discussed in Section~\ref{sec:single crystal}. The maximum strength of the samples having the same crystal orientations are approximately identical for all particle configurations. As an example, for the material with cube orientation (shown with point-dashed lines in Figs.~\ref{fig:Crystal_orientation_extracted}a and~b) the maximum strength is around $105$~MPa for $G_c^{\textrm{Al-M}}=4$ and $110$~MPa for $8$~Jm$^{-2}$. These values are higher for the material with random orientation (solid lines in Figs.~\ref{fig:Crystal_orientation_extracted}a and~b) with about $160$~MPa for $G_c^{\textrm{Al-M}}=4$ and $165$~MPa for $8$~Jm$^{-2}$, and even higher for the material with rotated-cube orientation (dashed lines in Figs.~\ref{fig:Crystal_orientation_extracted}a and~b) with about $180$~MPa for $G_c^{\textrm{Al-M}}=4$ and $185$~MPa for $8$~Jm$^{-2}$. 
The fracture strain obtained using $G_c^{\textrm{Al-M}}=4$~Jm$^{-2}$ (Fig.~\ref{fig:Crystal_orientation_extracted}) for the single crystal material with cube orientation demonstrates that the sample with particles parallel to the loading direction have the lowest value, similar to that of the random orientation. As already discussed in Section~\ref{sec:single crystal}, this was the case for debonding of the particles from the matrix, see Figs.~\ref{fig:Circle_to_ellipse_StressStrainGc4}b and~\ref{fig:Circle_to_ellipseGc48}b. 
However, for the case of the rotated-cube crystal, the fracture strain is approximately identical for all particle configurations. For a higher value of $G_c^{\textrm{Al-M}}=8$~Jm$^{-2}$ (Fig.~\ref{fig:Crystal_orientation_extracted}d), the fracture strain of the material with different particle configurations (circular, and ellipsoidal particles aligned parallel, 45$^\circ$ inclined, and perpendicular to the loading direction) for the material with rotated-cube orientation is almost unchanged, while for the single crystal with cube orientation, the results show a similar trend to that of using $G_c^{\textrm{Al-M}}=4$~Jm$^{-2}$, see the point-dashed lines in Figs.~\ref{fig:Crystal_orientation_extracted}c and~d. 
This finding reveals that increasing $G_c^{\textrm{Al-M}}$ from $4$ to $8$~Jm$^{-2}$ for the cube orientation does not change the damage mechanism and the single crystal with ellipsoidal particles aligned parallel to the loading direction fails first compared to the other cases with other particle configurations. 
A similar trend is observed for the magnitude of the fracture work, where the single crystal samples with cube orientation show a similar trend for both $G_c^{\textrm{Al-M}}=4$ and $8$~Jm$^{-2}$, see Figs.~\ref{fig:Crystal_orientation_extracted}e and~f. 
However, for the case of rotated-cube, the variation of the fracture work for different particle configurations is similar to the result of the random orientation for both values of $G_c^{\textrm{Al-M}}=4$ and $8$~Jm$^{-2}$, as shown by the dashed lines in Figs.~\ref{fig:Crystal_orientation_extracted}e and~f. 
It needs to be mentioned that for $G_c^{\textrm{Al-M}}=8$~Jm$^{-2}$, the increase of the fracture work for the parallel case is very small, as shown by the dashed line in Fig.~\ref{fig:Crystal_orientation_extracted}f. 
These results reveal that the change of the damage mechanism between particle/matrix debonding and particle cracking does depend on the matrix' crystal orientation. In a polycrystal, where different orientations are present in the microstructure, it is expected that the damage mechanism is a mixture of particle/matrix debonding in some grains, while particle breakage can prevail in others.

\clearpage
\subsection{Polycrystal}
\label{sec:polycrystal}
\begin{figure}
	\centering
	\includegraphics[width=0.9\linewidth]{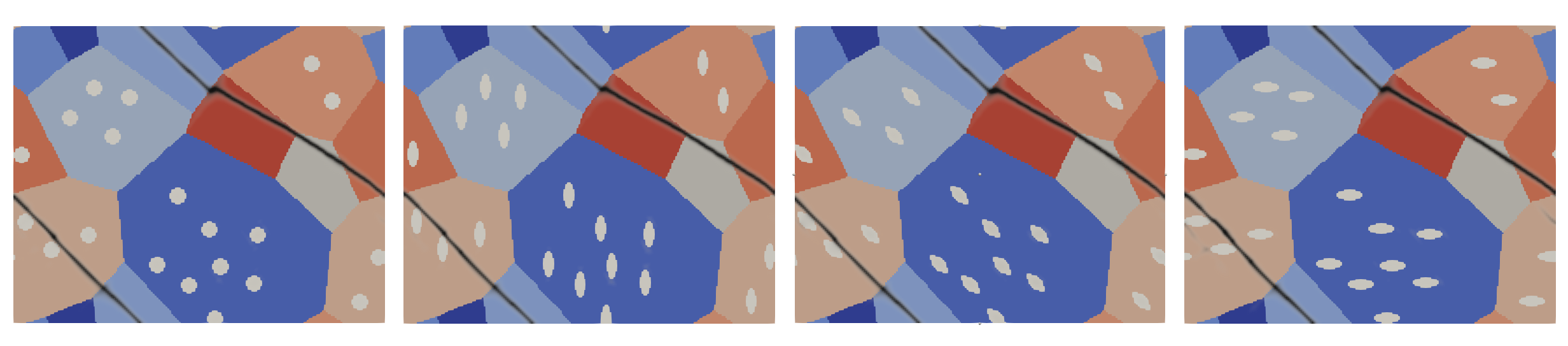}\\
	(a)\hspace*{0.21\linewidth}(b)\hspace*{0.21\linewidth}(c)\hspace*{0.21\linewidth}(d)
	\caption{Crack propagation path for four particle configurations (a-d) using $G_c^{\textrm{Al-M}}=10$~Jm$^{-2}$. The crack initiates from Mg$_2$Si particle/aluminum interface and propagates through the aluminum matrix. The particle critical energy release rate is given by $G_c^{\textrm{Mg$_2$Si}}=16$~Jm$^{-2}$.}
	\label{fig:Circle_to_ellipseGc10}
\end{figure}

\begin{figure}
	\centering
	\includegraphics[width=0.675\linewidth]{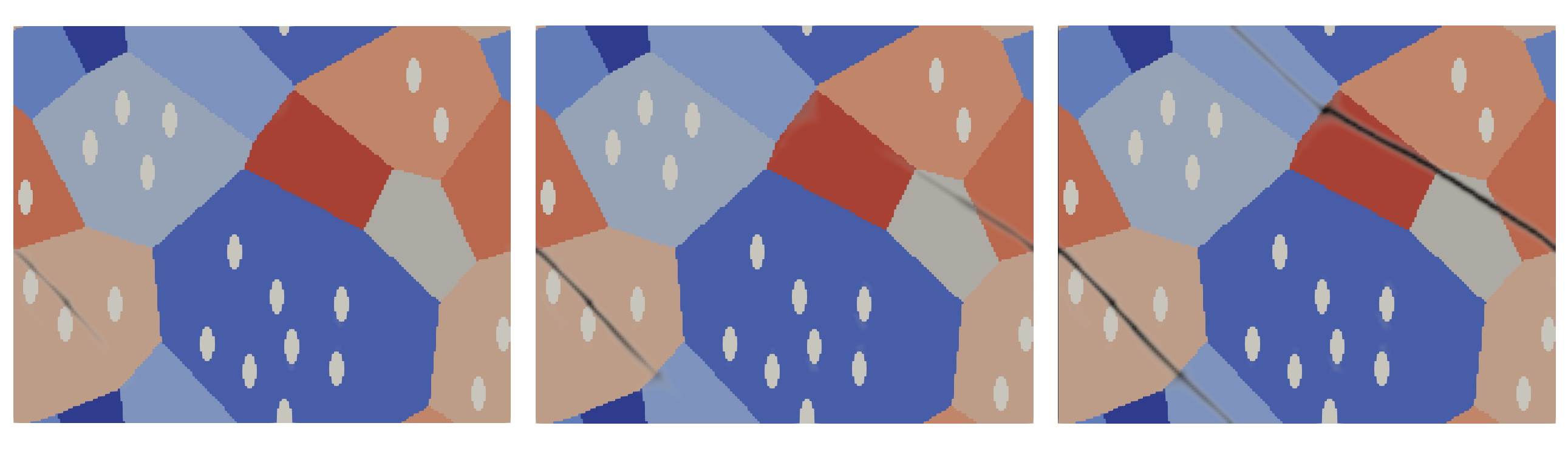}\\
	(a)\hspace*{0.21\linewidth}(b)\hspace*{0.21\linewidth}(c)
	\caption{Snapshots of the crack initiation and propagation (a-c) for the polycrystalline alloy with particle configurations shown in Fig.~\ref{fig:Circle_to_ellipseGc10} using $G_c^{\textrm{Al-M}}=10$~Jm$^{-2}$. The crack initiates from Mg$_2$Si particle/aluminum interface and propagates through the aluminum matrix.}
	\label{fig:CrackInitiation}
\end{figure}

\begin{figure}
	\centering
	\includegraphics[width=0.4\linewidth]{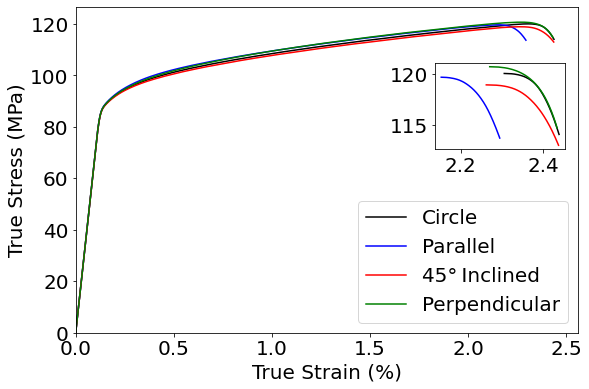}
	\includegraphics[width=0.4\linewidth]{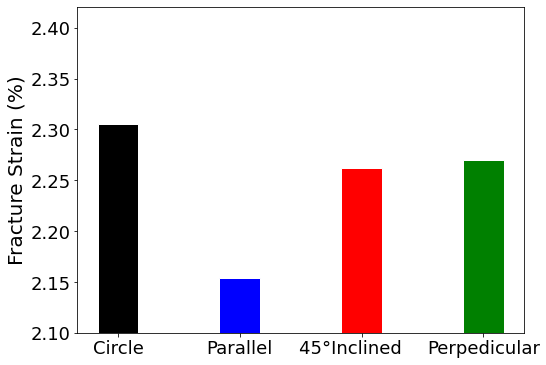}\\
	(a)\hspace*{0.4\linewidth}(b)\\
	\includegraphics[width=0.4\linewidth]{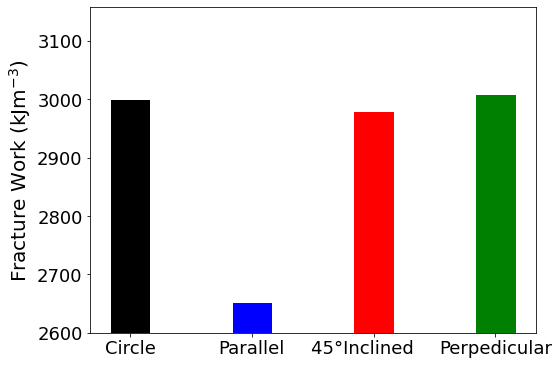}
	\includegraphics[width=0.4\linewidth]{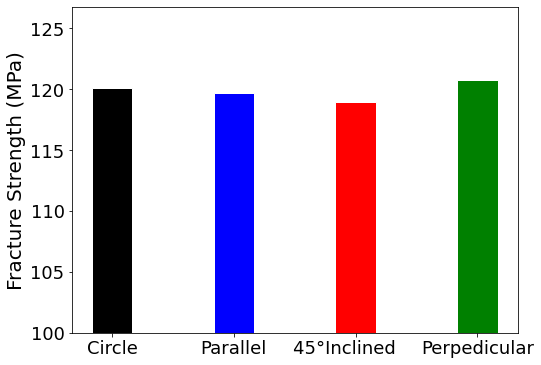}\\
	(c)\hspace*{0.4\linewidth}(d)
	\caption{(a) The stress-strain response of the uniaxial loading test for aluminum polycrystals with randomly distributed Mg$_2$Si particles inside, where $G_c^{\textrm{Al-M}}=10$~Jm$^{-2}$ and $G_c^{\textrm{Mg$_2$Si}}=16$~Jm$^{-2}$. The colors correspond to particles shape in case of circular ones and their orientation about the loading direction for ellipsoidal particles. The fracture strain (b), fracture work (b), and fracture strength (d) of the samples are extracted from the stress-strain curves.}
	\label{fig:CE_StressStrain_MultiGrain_Gc10}
\end{figure}

For a polycrystalline aluminum alloy with randomly distributed particles, similar particle configurations (circular and ellipsoidal Mg$_2$Si particles aligned parallel, 45$^\circ$ inclined, and perpendicular to the loading direction) are simulated. 
The simulation setup is similar as for the single crystals, see Section~\ref{sec:single crystal}. However, fewer particles are used to avoid overlap between particles and grain boundaries. The damaged microstructure of the polycrystal samples including circular and ellipsoidal particles aligned parallel, 45$^\circ$ inclined, and perpendicular to the loading direction are shown in Figs.~\ref{fig:Circle_to_ellipseGc10}a-d. When using $G_c^{\textrm{Al-M}}=10$~Jm$^{-2}$, which exceeds the maximum value used for the single crystals, the cracks initiate from the Mg$_2$Si/aluminum interfaces. To show the damage initiation and propagation, we have chosen a series of snapshots for the uniaxial loading case with ellipsoidal particles parallel to the loading direction, Fig.~\ref{fig:CrackInitiation}. The damage propagates through the matrix without considerable particle failure, i.e. fracture is entirely ductile in this case. The reason for this behavior, at a relatively high critical energy release rate of $G_c^{\textrm{Al-M}}=10$~Jm$^{-2}$, is explained by the strengthening provided by the presence of grain boundaries. In polycrystalline materials, grain boundaries act as barriers against dislocation motion, so that dislocations pile-up and enhance the strain hardening. Even though the constitutive model does not account for the Hall-Petch effect, we do observe additional strengthening due to the sole presence of grain boundaries. This is due to the additional compatibility constraints at the grain boundaries. As a result, the plastic energy for ductile fracture is reached at lower strains than in single crystals which undergo weaker strain hardening. Therefore, the crack initiates in the aluminum matrix while almost no damage is observed for the Mg$_2$Si particles, even not when using $G_c^{\textrm{Al-M}}=10$~Jm$^{-2}$, see Fig.~\ref{fig:Circle_to_ellipseGc10}. 
The stress-strain response of the polycrystals is shown in Fig.~\ref{fig:CE_StressStrain_MultiGrain_Gc10}a, where the colors correspond to the particles' shape (for circular particles) and their orientation to the loading direction (for ellipsoidal particles). 
The magnification in Fig.~\ref{fig:CE_StressStrain_MultiGrain_Gc10}a shows the region where fracture starts in each sample (maximum stress of the true stress-strain curves). The fracture strain, fracture work, and fracture strength extracted from the stress-strain curves are shown in Fig.~\ref{fig:CE_StressStrain_MultiGrain_Gc10}b to~d, respectively. Similar to the single crystal case using $G_c^{\textrm{Al-M}}=4$~Jm$^{-2}$, see Figs.~\ref{fig:Circle_to_ellipse_StressStrainGc4}, the polycrystalline material with ellipsoidal particles parallel to the load axis show the lowest fracture strain and fracture work (Fig.~\ref{fig:CE_StressStrain_MultiGrain_Gc10}b and~c). 
However, for the other model materials, the difference between the values of the fracture strain is only $0.05\%$, i.e. their damage response is approximately identical. 
Likewise, the fracture work for the other samples (circular and ellipsoidal particles aligned 45$^\circ$ inclined, and perpendicular to the loading direction) only varies slightly, Fig.~\ref{fig:CE_StressStrain_MultiGrain_Gc10}c. The fracture strength remains approximately identical for all particle configurations as well, see Fig.~\ref{fig:CE_StressStrain_MultiGrain_Gc10}d.

\begin{figure}
	\centering
	\includegraphics[width=0.4\linewidth]{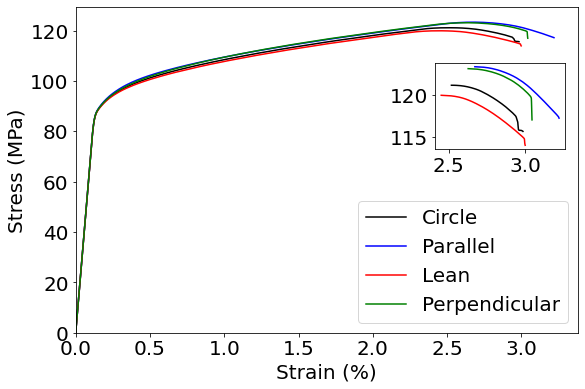}
	\includegraphics[width=0.4\linewidth]{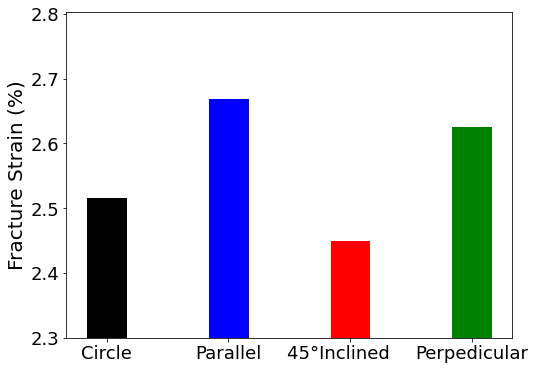}\\
	(a)\hspace*{0.4\linewidth}(b)\\
	\includegraphics[width=0.4\linewidth]{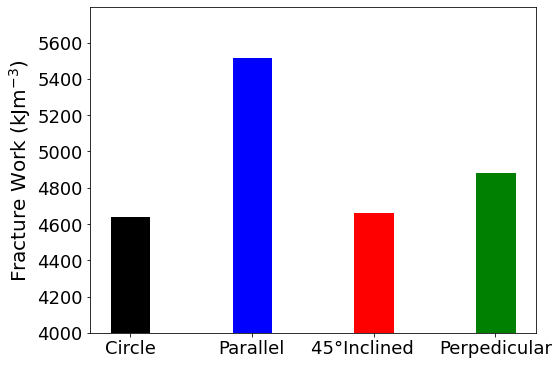}
	\includegraphics[width=0.4\linewidth]{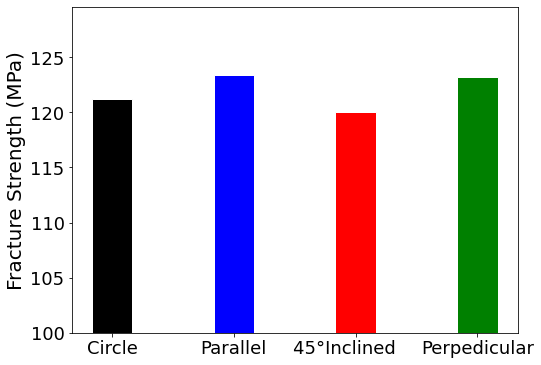}\\
	(c)\hspace*{0.4\linewidth}(d)\\
	\caption{(a) The stress-strain response of the uni-axial tensile test for polycrystalline aluminum alloys with randomly distributed Mg$_2$Si particles inside, where $G_c^{\textrm{Al-M}}=16$~Jm$^{-2}$ and $G_c^{\textrm{Mg$_2$Si}}=16$~Jm$^{-2}$. The colors correspond to particles shape in case of circular ones and their orientation about the loading direction for ellipse shape particles. The fracture strain (b), fracture work (b), and fracture strength (d) of the samples are extracted from the stress-strain curves in (a).}
	\label{fig:CE_StressStrain_MultiGrain_Gc16}
\end{figure}

The results obtained for polycrystal simulations with a higher value for the critical energy release rate $G_c^{\textrm{Al-M}}=16$~Jm$^{-2}$ are depicted in Fig.~\ref{fig:CE_StressStrain_MultiGrain_Gc16}, where Fig.~\ref{fig:CE_StressStrain_MultiGrain_Gc16}a represents the stress-strain curves, and Fig.~\ref{fig:CE_StressStrain_MultiGrain_Gc16}b-d show the fracture strain, fracture work, and fracture strength of these simulations, respectively. 
The model material with ellipsoidal particles aligned parallel to the loading direction shows the highest fracture strain and fracture work (Figs.~\ref{fig:CE_StressStrain_MultiGrain_Gc16}b and~c) among all polycrystals, similar as observed for the single crystal cases using $G_c^{\textrm{Al-M}}=8$~Jm$^{-2}$, see Fig.~\ref{fig:Circle_to_ellipse_StressStrainGc8}. 

\begin{figure}
	\centering
	\includegraphics[width=0.9\linewidth]{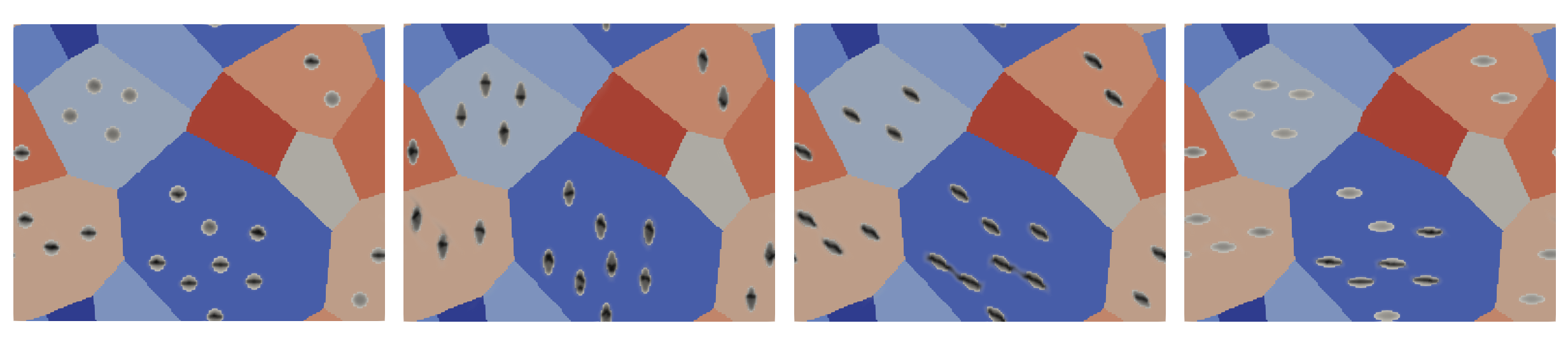}\\
	(a)\hspace*{0.21\linewidth}(b)\hspace*{0.21\linewidth}(c)\hspace*{0.21\linewidth}(d)
	\caption{Cracking of particles for four particle configurations (a-d) using $G_c^{\textrm{Al-M}}=16$~Jm$^{-2}$. Crack initiates by breaking of Mg$_2$Si particles and propagates through the aluminum matrix. The particle critical energy release rate is given by $G_c^{\textrm{Mg$_2$Si}}=16$~Jm$^{-2}$.}
	\label{fig:Circle_to_ellipseGc16}
\end{figure}

The damage morphologies of the polycrystalline microstructures are shown in Fig.~\ref{fig:Circle_to_ellipseGc16} for the four particle configurations. 
Cracking starts from by breaking of the Mg$_2$Si precipitates and subsequent crack propagation into the adjacent aluminum matrix. An example is shown in Fig.~\ref{fig:Circle_to_ellipseGc16}c.

The simulation of two different types of damage initiation scenarios in this work, debonding of particles from the matrix and their breaking, reveals that differences in topology, shape, spacing and orientation of the particles in aluminum alloys can substantially affect the materials' mechanical properties, particularly their damage mode. 
These different types of failure modes have been indeed observed in the alloy AA 6060 \cite{Lassance2007}. For a tensile test at room temperature, it was found that the particles break, while at an elevated test temperature of 550 $^\circ$C they debond from the aluminum matrix. Furthermore, Babout et al. \cite{Babout2004} have found two different types of damage mechanisms by using two different aluminum matrices, namely pure Al and Al2124, reinforced with spherical hard ceramic particles. They observed that for commercially pure aluminum, particles debond from the matrix, while they break for Al2124 alloy. The results of the simulations shown in Figs.~\ref{fig:Circle_to_ellipseGc10} and~\ref{fig:Circle_to_ellipseGc16} qualitatively describe these two regimes, i.e. particle/matrix debonding and particle breakage, respectively, as observed in \cite{Lassance2007,Babout2004}.

\section{Conclusions}
In this paper, a new ductile fracture model is formulated and applied to investigate the particle- and microstructure-dependent initiation and evolution of damage in aluminum alloys. 
For an aluminum alloy with micro-scale particles (Mg$_2$Si precipitates), a ductile fracture model is used for describing the response of the aluminum matrix, while a brittle fracture model is used for the particles. 
The effect of the particles' shapes and orientations (relative to the loading axis) is studied by simulating both, single crystal and polycrystal model materials with circular and ellipsoidal particles (varying their longitudinal axis relative to the loading direction). It is demonstrated that depending on the aluminum matrix's critical energy release rate, damage can initiate either from the particle/matrix interface (debonding) or by particle cracking. 
Based on the type of damage initiation, different particle topologies lead to differences in damage behavior. For particle/matrix damage initiation, circular and ellipsoidal particles arranged perpendicular to the loading direction leads to the highest ductility and fracture work, while for particle breaking, model materials with parallel ellipsoidal particles show the best results (i.e. highest damage resistance). For the case of single crystals, the change in damage mechanism depends on the crystal's orientation.




\bibliography{mybibfile}


\end{document}